\documentclass[reprint,amssymb,amsmath,amsfonts,aps,floatfix,showpacs]{revtex4-1}
\usepackage{graphicx, units, xspace} 

\newcommand{\nv}{\ensuremath{\vec{0}}\xspace}
\newcommand{\defi}{\mathrel{\mathop:}=}

\newcommand{\kl}[1]{\mathopen{}\mathclose\bgroup\left(#1\aftergroup\egroup\right)}
\newcommand{\klg}[1]{\mathopen{}\mathclose\bgroup\left\{#1\aftergroup\egroup\right\}}
\newcommand{\abs}[1]{\mathopen{}\mathclose\bgroup\left| #1 \aftergroup\egroup\right|}
\newcommand{\E}[1]{\cdot 10^{#1}}
\DeclareMathOperator{\vspan}{span}
\DeclareMathOperator{\mdet}{det}

\begin{document}

\title{Natural units and the vector space of physical values}
\author{Gerrit Ansmann}
\affiliation{University of Bonn, Germany}

\begin{abstract}
We explore the mathematical foundations of the vector space of physical dimensions introduced in \emph{A.~Maksymowicz, Am. J. Phys. 44\,(3), 1976,} and extend this formalism to the vector space of physical values.
As different unit systems correspond to different bases of this vector space, our formalism may find use for introducing the concept of natural units and transforming physical values between unit systems.
\end{abstract}


\maketitle

\section{Introduction}
The term \textit{natural units} usually refers to the combination of the following two techniques:
employing a different set of base units, usually natural constants relevant to the respective field \cite{Planck,Bartlett}; and setting some or all of these base units to unity, i.e., omitting them in calculations \cite{Bartlett,Desloge}.
Since these techniques are usually introduced together, at times with little more than ``\(\hbar=c=1\)'', students may be confused about their validity and how to apply them.
We here try to address this issue by introducing a mathematical formalism for unit systems and transformations between them which only builds on basic linear algebra.

To this purpose, we incorporate numerical values into the dimension space introduced in Ref.~\cite{Maksymowicz} and thereby extend it to the vector space of physical values.
In this vector space (as in the dimension space), transformations between unit systems correspond to basis transformations.
In addition, we explore the formalism's mathematical background and formally introduce this vector space on an axiomatic level.
Finally, we discuss the possible merit of our formalism for introducing the concept of natural units as well as some fundamental concepts of linear algebra.

\section{The vector space of physical values}

Some general remarks on notation:
Though the aforementioned technique of setting some or all base units to unity can easily complement our formalism, we here refrain from using it for clarity's sake.
We denote vector operations with encircled symbols (\(\oplus\), \(\ominus\) and \(\odot\)) to avoid confusion with the corresponding regular aoperations.
\(\nv\)~denotes the null vector.
\textit{Physical value} or just \textit{value} refers to a possible value of a dimensionful or dimensionless, scalar physical quantity~\cite{VIM}.
Examples for physical values are \unit[1]{fm}, \unit[197]{MeV}, \(4.2\), \(\pi\)  or the value of \(\hbar\).
A physical value can be represented in different ways, e.g., \unit[4.3]{kW} and \unitfrac[4300]{J}{s} are two representations of the same physical value.

Let \(\Omega\) denote the set of all positively signed physical values.
We regard \(\Omega\) as an \(\mathbb{R}\)-vector space, with the multiplication of physical values as vector addition and their exponentiation as scalar multiplication, i.e.:
\begin{align*}
	x \oplus y &\defi x y &\forall x,y \in \Omega\\
	\alpha \odot x &\defi x^\alpha &\forall x \in \Omega, \alpha \in \mathbb{R}
\end{align*}
For example, \(\unit[4.2]{m s^{-2}}\) written with these operations is \(4.2 \oplus \kl{ \unit{m} \oplus \kl{\kl{-2} \odot \unit{s}}}\) and \(\unit[2]{N}\cdot\unit[3]{m}\) is \(\kl{2\oplus\unit{N}}\oplus\kl{3 \oplus \unit{m}}\).

To make \(\Omega\) closed under \(\odot\), we have to accept values such as \(\pi \odot \unit{kg} = \unit{kg^\pi}\) or \(\frac{1}{2} \odot \unit{m} = \unit{\sqrt{m}}\).
However, as we will discuss in Sec.~\ref{avoiding}, whether these come up in application is tied to the unit system and not a consequence of our formalism, and they can be (and usually are) avoided by choosing the unit system accordingly.

We will now reason that the vector-space axioms hold for \(\Omega\) with \(\oplus\) and \(\odot\) and explore some of its fundamental properties in the process:
\begin{itemize}
	\item Commutativity and associativity of \(\oplus\) are inherited from the multiplication of physical values:
	\begin{align*}
	x \oplus y = x y &= y x = y \oplus x\\
	x \oplus \kl{y \oplus z} = x \kl{ y z} &= \kl{ x y } z = \kl{ x \oplus y } \oplus z \\ &\quad\forall x,y,z \in \Omega
	\end{align*}
	\item Since \(1 \oplus x = 1x = x\) for all \(x \in \Omega\), the neutral element of \(\oplus\) is \(1\in\Omega\), i.e., \(\nv=1\).
	\item Since \(x \oplus \kl{x^{-1}} = x x^{-1} = 1 = \nv\) for all \(x \in \Omega\), the inverse of \(x\) under \(\oplus\) is \(x^{-1} \in \Omega\), i.e., \(\ominus x = x^{-1}\).
	\item The real number 1 is the identity of scalar multiplication: \(1 \odot x = x^1 = x \;\forall x \in \Omega\).
	\item Compatibility of multiplications:
	\displaywidth=\linewidth
	\displayindent=-\leftskip
	\begin{align*}
	\kl{\alpha \beta} \odot x = x^{\alpha \beta} &= \kl{x^\beta}^\alpha = \alpha \odot \kl{ \beta \odot x}\\ &\quad\forall x \in \Omega; \alpha, \beta \in \mathbb{R}
	\end{align*}
	\item Distributivities:
	\begin{align*}
		\kl{\alpha + \beta} \odot x = x^{\alpha + \beta} &= x^\alpha x^\beta = \alpha \odot x \oplus  \beta \odot x\\
		\alpha \odot \kl{ x \oplus y} = \kl{x y}^\alpha &= x^\alpha y^\alpha = \alpha \odot x \oplus \alpha \odot y \\
		&\quad \forall x,y \in \Omega; \alpha,\beta \in \mathbb{R}
	\end{align*}
\end{itemize}

\section{Different unit systems and transforming between them}

As the SI~base units together with the scalar~10 are independent by means of multiplication and exponentiation, they are linearly independent and thus form indeed an algebraic basis of their span.
For simplicity, we restrict the discussion to the subspace spanned by \(\mathcal{S} \defi \klg{10, \unit{m}, \unit{s}, \unit{kg}}\) in the following examples.
Using~\(\mathcal{S}\) as an ordered basis, we can write for example and later use:
\begin{equation}
\begin{aligned}
	10 &= 10^1 = 1 \odot 10 = \kl{1,0,0,0}_\mathcal{S}\\
	c
	&\approx \unitfrac[3 \E{8}]{m}{s}
	= \unitfrac[10^{\log\kl{3} + 8}]{m}{s}
	\approx 10^{8.5}\, \unit{m \, s^{-1}}\\
	&= 8.5 \odot 10 \oplus \unit{m} \oplus \kl{-1} \odot \unit{s}
	= \kl{8.5,\; 1, -1,\; 0}_\mathcal{S}  \\
	\hbar
	&\approx \unit[1.05 \E{-34.0}]{J s}
	= \unit[10^{\log\kl{1.05} -34.0}]{J s}\\
	&\approx \kl{-34.0} \odot 10 \oplus 2 \odot \unit{m} \oplus \kl{-2} \odot \unit{s} \oplus \unit{kg} \\
	&= \kl{-34.0,\; 2, -1,\; 1}_\mathcal{S}  \\
	\unit[1]{MeV}
	&\approx \unit[1.6 \E{-13}]{J}
	= \unit[10^{\log\kl{1.6}-13}]{J}\\
	&\approx \kl{-12.8} \odot 10 \oplus 2 \odot \unit{m} \oplus \kl{-2} \odot \unit{s} \oplus \unit{kg} \\
	&= \kl{-12.8,\; 2, -2,\; 1}_\mathcal{S} ,
 \end{aligned}
 \label{eq:representation}
\end{equation}
where the subscript indicates the basis of the component representation.

In our formalism, different unit systems do correspond to different bases.
For example, one common system of (partially) natural units used by high-energy physics corresponds to the basis \(\mathcal{N} = \klg{10, c, \hbar, \unit[1]{MeV}}\) of \(\vspan\kl{\mathcal{S}}\).
Using  Eq.~\ref{eq:representation}, we obtain basis transformation matrices from \(\mathcal{N}\) to \(\mathcal{S}\) and vice versa:

\begin{align*}
	T^\mathcal{N}_\mathcal{S}
	&\approx \kl{\begin{array}{rrrr}
		1 & 8.5 & -34.0 & -12.8 \\
		0 & 1 & 2 & 2 \\
		0 & -1 & -1 & -2 \\
		0 & 0 & 1 & 1
	\end{array}};
	\\
	T_\mathcal{N}^\mathcal{S}
	= \kl{T^\mathcal{N}_\mathcal{S}}^{-1}
	&\approx \kl{\begin{array}{rrrr}
		1 & 12.7 & 21.2 & 29.7 \\
		0 & 1 & 0 & -2 \\
		0 & 1 & 1 & 0 \\
		0 & -1 & -1 & 1
	\end{array}}.
\end{align*}
With this, we can for example transform \unit[1]{fm} to the above natural units, i.e., the basis~\(\mathcal{N}\) (using \(T_\mathcal{S}^\mathcal{N} \kl{-15,1,0,0} \approx \kl{-2.3,1,1,-1} \)):

\begin{align*}
 \unit[1]{fm}
&= \kl{-15,1,0,0}_\mathcal{S}
\approx \kl{-2.3,1,1,-1}_\mathcal{N}\\
&= 10^{-2.3} \frac{\hbar c}{\unit{MeV}}
\approx \frac{\hbar c}{\unit[197]{MeV}}
\end{align*}

\subsection{Avoiding non-integer powers of units}\label{avoiding}

A usual requirement for a unit system is that physically meaningful values can be represented without any unit being raised to a non-integer power.
For example one would avoid using \(\unit{qm}\defi\unit{m^2}\) as a base unit, as this would render the unit of length to be \(\sqrt{\unit{qm}}\).
Using our formalism and assuming that at least one basis element of a unit system is a positive number (usually~10), we can classify those unit systems for which the above requirement is fulfilled:
Suppose we have an established unit system corresponding to the basis \(\mathcal{A} \defi \klg{a_0,a_1,\ldots,a_n}\) with \(a_0 \in \mathbb{R}_+\) and \(n \in \mathbb {N}\) and this system fulfils the above requirement.
Suppose further that we want to introduce a new unit system corresponding to the basis \(\mathcal{B} \defi \klg{b_0, b_1,\ldots, b_n}\) with \(b_0 \in \mathbb{R}_+\) and \(\vspan\kl{\mathcal{B}} = \vspan\kl{\mathcal{A}}\).
We can then write the transformation matrix \(T_\mathcal{A}^\mathcal{B}\) from \(\mathcal{B}\) to \(\mathcal{A}\) and its inverse as follows:
\begin{align*}
	T^\mathcal{B}_\mathcal{A}
	&= \kl{\begin{array}{c|ccc}
 		\star		& \star	& \cdots	& \star	\\\hline
		0		&	&			&	\\
		\vdots	&	& \hat{T}^\mathcal{B}_\mathcal{A}	&	\\
		0		&	&			&	
	\end{array}};\\
	T_\mathcal{B}^\mathcal{A}
	= \kl{T^\mathcal{B}_\mathcal{A}}^{-1}
		&= \kl{\begin{array}{c|ccc}
		\star		& \star	& \cdots	& \star	\\\hline
		0		&	&			&	\\
		\vdots	&	& \kl{\hat{T}^\mathcal{B}_\mathcal{A}}^{-1}	&	\\
		0		&	&			&	
	\end{array}},
\end{align*}
where \(\star\) represents some real number.
The leftmost columns only contain non-zero entries in the first row because \(a_0 = \kl{\log\kl{a_0}/\log\kl{b_0}} \odot b_0\).

As \(\mathcal{A}\) fulfils the requirement, each~\(b_i\) for \(i \geq 1\) can be expressed in units from~\(\mathcal{A}\) without raising any \(a_i\) for \(i \geq 1\) to a non-integer power and thus \(\hat{T}^\mathcal{B}_\mathcal{A}\)~has only integer elements.
Analogously, \(\mathcal{B}\) fulfils the requirement if and only if \(\kl{\hat{T}^\mathcal{B}_\mathcal{A}}^{-1}\) has integer elements only.
If \(\hat{T}^\mathcal{B}_\mathcal{A}\) having only integer elements is given, \(\kl{\hat{T}^\mathcal{B}_\mathcal{A}}^{-1}\) having only integer elements is equivalent to \(\abs{\mdet\kl{\hat{T}^\mathcal{B}_\mathcal{A}}}=1\) and to \(\hat{T}^\mathcal{B}_\mathcal{A}\) being a unimodal matrix.

\(\hat{T}^\mathcal{B}_\mathcal{A}\) and \(\kl{\hat{T}^\mathcal{B}_\mathcal{A}}^{-1}\) correspond to transformation matrices in the vector space (or more precisely, the \(\mathbb{Z}\)\nobreakdash-module~\footnote{A \(\mathbb{Z}\)-module is an algebraic structure that complies with the same axioms as a vector space, but with the scalars coming only from the ring of integers~\(\mathbb{Z}\) instead of a field.}) of physical dimensions introduced in Ref.~\cite{Maksymowicz}.

\section{Conclusions}

We introduced the vector space of physical values and thereby provided a mathematical foundation for unit systems and transformations between them.
From a practical perspective, our formalism allows transforming a representation of a physical value to a different unit system with one matrix--vector multiplication, using the same matrix for a given pair of unit systems.

In particular, we hope that our formalism has didactic values for introducing natural units as it not only addresses how to apply this technique but substantiates its mathematical foundations and validity.
Though our formalism only requires basic concepts from linear algebra, it may be challenging for some students due its unusual vector addition and scalar multiplication.
On the other hand, the latter may be beneficial if our vector space is used as an example when teaching linear algebra, as it may help to illustrate how abstract, general and diverse the presented concepts are and act prophylactic against students conflating vector operations and regular operations.
Furthermore, using transformations between unit systems as example for basis transformations may help overcoming certain difficulties as there is no straightforward canonical basis.
Finally, our approach exemplifies the value of axiomatic systems, as validating a few axioms was all we had to do to be able to safely apply well-known concepts from linear algebra to a new structure.

\section*{Acknowledgements}

I am grateful to N.~Ansmann, S.~Bialonski and J.~Wilting for critical comments on earlier versions of the manuscript.


\begin{thebibliography}{6}%
\makeatletter
\providecommand \@ifxundefined [1]{%
 \@ifx{#1\undefined}
}%
\providecommand \@ifnum [1]{%
 \ifnum #1\expandafter \@firstoftwo
 \else \expandafter \@secondoftwo
 \fi
}%
\providecommand \@ifx [1]{%
 \ifx #1\expandafter \@firstoftwo
 \else \expandafter \@secondoftwo
 \fi
}%
\providecommand \natexlab [1]{#1}%
\providecommand \enquote  [1]{``#1''}%
\providecommand \bibnamefont  [1]{#1}%
\providecommand \bibfnamefont [1]{#1}%
\providecommand \citenamefont [1]{#1}%
\providecommand \href@noop [0]{\@secondoftwo}%
\providecommand \href [0]{\begingroup \@sanitize@url \@href}%
\providecommand \@href[1]{\@@startlink{#1}\@@href}%
\providecommand \@@href[1]{\endgroup#1\@@endlink}%
\providecommand \@sanitize@url [0]{\catcode `\\12\catcode `\$12\catcode
  `\&12\catcode `\#12\catcode `\^12\catcode `\_12\catcode `\%12\relax}%
\providecommand \@@startlink[1]{}%
\providecommand \@@endlink[0]{}%
\providecommand \url  [0]{\begingroup\@sanitize@url \@url }%
\providecommand \@url [1]{\endgroup\@href {#1}{\urlprefix }}%
\providecommand \urlprefix  [0]{URL }%
\providecommand \Eprint [0]{\href }%
\providecommand \doibase [0]{http://dx.doi.org/}%
\providecommand \selectlanguage [0]{\@gobble}%
\providecommand \bibinfo  [0]{\@secondoftwo}%
\providecommand \bibfield  [0]{\@secondoftwo}%
\providecommand \translation [1]{[#1]}%
\providecommand \BibitemOpen [0]{}%
\providecommand \bibitemStop [0]{}%
\providecommand \bibitemNoStop [0]{.\EOS\space}%
\providecommand \EOS [0]{\spacefactor3000\relax}%
\providecommand \BibitemShut  [1]{\csname bibitem#1\endcsname}%
\let\auto@bib@innerbib\@empty
\bibitem [{\citenamefont {Planck}(1900)}]{Planck}%
  \BibitemOpen
  \bibfield  {author} {\bibinfo {author} {\bibfnamefont {M.}~\bibnamefont
  {Planck}},\ }\href {\doibase 10.1002/andp.19003060105} {\bibfield  {journal}
  {\bibinfo  {journal} {Annalen der Physik}\ }\textbf {\bibinfo {volume}
  {306}},\ \bibinfo {pages} {69} (\bibinfo {year} {1900})}\BibitemShut
  {NoStop}%
\bibitem [{\citenamefont {Bartlett}(1974)}]{Bartlett}%
  \BibitemOpen
  \bibfield  {author} {\bibinfo {author} {\bibfnamefont {D.~F.}\ \bibnamefont
  {Bartlett}},\ }\href {\doibase 10.1119/1.1987631} {\bibfield  {journal}
  {\bibinfo  {journal} {Am. J. Phys.}\ }\textbf {\bibinfo {volume} {42}},\
  \bibinfo {pages} {148} (\bibinfo {year} {1974})}\BibitemShut {NoStop}%
\bibitem [{\citenamefont {Desloge}(1994)}]{Desloge}%
  \BibitemOpen
  \bibfield  {author} {\bibinfo {author} {\bibfnamefont {E.~A.}\ \bibnamefont
  {Desloge}},\ }\href {\doibase 10.1119/1.17599} {\bibfield  {journal}
  {\bibinfo  {journal} {Am. J. Phys.}\ }\textbf {\bibinfo {volume} {62}},\
  \bibinfo {pages} {216} (\bibinfo {year} {1994})}\BibitemShut {NoStop}%
\bibitem [{\citenamefont {Maksymowicz}(1976)}]{Maksymowicz}%
  \BibitemOpen
  \bibfield  {author} {\bibinfo {author} {\bibfnamefont {A.}~\bibnamefont
  {Maksymowicz}},\ }\href {\doibase 10.1119/1.10470} {\bibfield  {journal}
  {\bibinfo  {journal} {Am. J. Phys.}\ }\textbf {\bibinfo {volume} {44}},\
  \bibinfo {pages} {295} (\bibinfo {year} {1976})}\BibitemShut {NoStop}%
\bibitem [{\citenamefont {{Joint Committee for Guides in
  Metrology}}(2008)}]{VIM}%
  \BibitemOpen
  \bibfield  {author} {\bibinfo {author} {\bibnamefont {{Joint Committee for
  Guides in Metrology}}},\ }\href {http://www.bipm.org/vim} {\emph {\bibinfo
  {title} International Vocabulary of Metrology}},\ \bibinfo {edition} {3\textsuperscript{rd}}\ ed. (\bibinfo {year}
  {2008})\BibitemShut {NoStop}%
\bibitem [{Note1()}]{Note1}%
  \BibitemOpen
  \bibinfo {note} {A \(\protect \mathbb {Z}\)-module is an algebraic structure
  that complies with the same axioms as a vector space, but with the scalars
  coming only from the ring of integers~\(\protect \mathbb {Z}\) instead of a
  field.}\BibitemShut {Stop}%
\end{thebibliography}
\end{document}